\documentclass[10pt,twocolumn,journal,shortpaper,compsoc]{IEEEtran}

\ifCLASSOPTIONcompsoc
  \usepackage[nocompress]{cite}
\else
  \usepackage{cite}
\fi

\ifCLASSINFOpdf
\else
\fi

\usepackage{amsmath}
\usepackage{amsfonts}
\usepackage{amssymb}
\usepackage{comment}
\usepackage{graphicx}
\usepackage{caption}
\usepackage{url}
\usepackage{hyperref}
\usepackage{booktabs}
\usepackage[table,xcdraw]{xcolor}
\usepackage[export]{adjustbox}
\usepackage{url}
\usepackage{mathtools}
\usepackage[]{mdframed}

\DeclareCaptionType{mycapequ}[][List of equations]
\newcommand{\argminF}{\mathop{\mathrm{argmin}}\limits}   %

\begin{document}
\title{A Comment on ``e-PoS: Making PoS Decentralized and Fair" }

\author{Suhyeon~Lee
        and~Seungjoo~Kim%
\IEEEcompsocitemizethanks{\IEEEcompsocthanksitem Suhyeon Lee is with Tokamak Network, Singapore and Korea University, Seoul, Korea.
\IEEEcompsocthanksitem Seungjoo Kim is with School of Cyber Security, Korea University, Seoul, Korea.\protect\\
E-mail: \{orion-alpha, skim71\}@korea.ac.kr }
}

\IEEEtitleabstractindextext{%
\begin{abstract}

Proof-of-Stake (PoS) is a prominent Sybil control mechanism for blockchain-based systems. In ``e-PoS: Making PoS Decentralized and Fair," Saad et al. (TPDS'21) introduced a new Proof-of-Stake protocol, e-PoS, to enhance PoS applications' decentralization and fairness. In this comment paper, we address a misunderstanding in the work of Saad et al. The conventional Proof-of-Stake model that causes the fairness problem does not align with the general concept of Proof-of-Stake nor the Proof-of-Stake cryptocurrencies mentioned in their paper.

\end{abstract}

}

\maketitle

\IEEEdisplaynontitleabstractindextext

\IEEEpeerreviewmaketitle

\IEEEraisesectionheading{\section{Introduction}\label{sec:introduction} }

\IEEEPARstart{C}{onsensus} mechanisms in blockchain prevent Sybil attacks but are also related to fairness and decentralization.
Among them, Proof-of-Stake (PoS) is a notable mechanism in the blockchain industry since it provides economically and environmentally beneficial aspects. It is based on a deposit made by miners to pick a block proposer. Staking is the act of securing tokens as a deposit.
As a result, a number of cryptocurrencies, like Ethereum, Cardano, Tezos, and Blackcoin, have included PoS in their Sybil control mechanism. 
In 2021, M. Saad et al. \cite{saad2021pos} proposed a new PoS scheme, named e-PoS, as a solution to decentralization and fairness issues of the conventional PoS in their paper titled ``e-PoS: Making PoS Decentralized and Fair''.

In this comment paper, we present that
the description of PoS in \cite{saad2021pos} does not match the general definition used in PoS-based coins. Thus, we conclude that the issues of decentralization and fairness they seek to address do not exist in the conventional PoS.

\section{Misunderstanding of PoS Leading to a False Questioning}

In this section, we demonstrate that the conventional PoS concept in Saad et al. is significantly different from real-world cryptocurrencies including PoS-based cryptocurrencies mentioned in their paper.

\subsection{Approach to Conventional PoS Concept}

In Saad et al., the authors attempted to address two main issues with the conventional PoS, namely, fairness and decentralization. For the conventional PoS, they mentioned Blackcoin and Nxt as PoS-based cryptocurrencies. 
We regard the conventional PoS to be the PoS concept which is already in use for real-world PoS applications and which is shared among these PoS applications. 
Therefore, we propose to refer to additional real-world cryptocurrencies to understand the conventional concept of PoS.

We analyze 6 PoS-based cryptocurrencies including 2 cryptocurrencies (Nxt and Blackcoin) mentioned in \cite{saad2021pos}. 
We selected Cosmos (ATOM) \cite{kwon2014tendermint}, Tezos (XTZ) \cite{PoSTezos}, Cardano (ADA) \cite{kiayias2017ouroboros}, and Algorand (ALGO) \cite{chen2018algorand} without loss of generality. 
Table \ref{table: comparison} shows the search results in \textit{Google Scholar} and market ranks of PoS cryptocurrencies. We searched the full name of each cryptocurrency with blockchain in quotation marks to avoid irrelevant results. For example, we searched ``Algorand'', and ``blockchain'' with the \textit{AND} condition.
The PoS cryptocurrencies that we introduced in this paper show equal or higher market ranks \footnote{\url{https://coinmarketcap.com/} May 3, 2022} and citations. We could find the common probability concept in these PoS applications.

\begin{table}[!htb]
\caption{Proof-of-Stake cryptocurrencies with search results and market ranks}
\label{table: comparison}
\centering
\begin{tabular}{c
>{\columncolor[HTML]{EFEFEF}}c 
>{\columncolor[HTML]{EFEFEF}}c cccc}
\toprule
Coin   & BLK  & NXT & ATOM & ADA & XTZ & ALGO \\ \toprule
\# of Search & 638  & 1710 & 2070  & 2640 & 1540 & 1890 \\ \midrule
Market Rank & 1486 & 1020 & 26   & 9   & 42  & 28   \\ \toprule
\end{tabular}
\end{table}

    \subsection{Comparison of the PoS next-block probability in Saad et al. and real-world PoS cryptocurrencies}

First of all, the description of the conventional PoS in \cite{saad2021pos} is the basis of their problem statement and evaluation. It is clearly given in the equation in Saad et al. (Eq. \ref{equation: next block in saad}) where $\alpha$ is a miner's stake, and $\beta$ is the total amount of stake in the blockchain. According to the equation, it is similar to an auction system. A 51\% attacker is always selected as a block proposer in the conventional PoS. It is the main reason which brings unfairness and centralization issues in the conventional PoS \cite{saad2021pos}.

Second, based on the conventional PoS concept, we provide the revised equation (Eq. \ref{equation: next block in real}). The miner's probability of being a block proposer is proportional in the range $ [0, 1]$ in Eq. \ref{equation: next block in real}. It always provides an opportunity to be a block proposer for a miner in real-world PoS cryptocurrencies while a 51\% miner is always a block proposer in Eq. \ref{equation: next block in saad}. 

\begin{center}
\begin{tabular}{|p{0.9\linewidth}|}\hline %
\rule{0pt}{1ex}%

\textbf{Equation in Saad et al.}. The probability to mint a next block in the conventional PoS

\begin{equation} \label{equation: next block in saad}
    Pr(\alpha, \beta) = \left \{ \begin{array}{lr}
        \alpha/\beta & \text{, $\alpha/\beta < 0.5$}  \\
        1 & \text{, $\alpha/\beta \geq 0.5$} 
    \end{array}  \right \} 
\end{equation} \\\hline

\end{tabular}
\end{center}

\begin{center}
\begin{tabular}{|p{0.9\linewidth}|}\hline %
\rule{0pt}{1ex}%

\textbf{Revised Equation}. The probability to mint a next block in the conventional PoS

\begin{equation} \label{equation: next block in real}
    Pr(\alpha, \beta) = \alpha/\beta 
\end{equation} 
\\ \hline
\end{tabular}
\end{center}

\begin{table}[!tb]
\caption{PoS leader selection processes and their probabilities under attackers can distribute their stakes into the minimum scale for the leader election}
\label{table: PoS comparison}
\begin{tabular}{lll}
\toprule
\multicolumn{1}{c}{PoS} & \multicolumn{1}{c}{Leader selection}                           & Probability             \\ \toprule
Peercoin \cite{PeercoinPoSCode}                & $\textsf{hash}(M_i, T) < D \times C \times A$       & $p_i = \frac{s_i a_i}{\sum^{n}_{k=1} s_k a_k}$ \\ \midrule
Blackcoin \cite{blackcoin}              & $\textsf{hash}(M_i, T) < D \times C$                   & $p_i = \frac{s_i}{\sum^{n}_{k=1} s_k}$     \\ \midrule
Nxt \cite{NxtWhitepaper}                    & $\textsf{hash}(M_i, T) < D \times C$                   & $p_i = \frac{s_i}{\sum^{n}_{k=1} s_k}$       \\ \midrule
Ouroborous \cite{kiayias2017ouroboros}             & $\text{leader} = \mathcal{F}(S, M)$       & $p_i = \frac{s_i}{\sum^{n}_{k=1} s_k}$       \\ \midrule
Algorand \cite{chen2018algorand}               & $\text{leader} = \argminF_{\text{participant } k} H(M_k)$  & $p_i = \frac{s_i}{\sum^{n}_{k=1} s_k}$     \\ \bottomrule
\end{tabular}
\end{table}

To get this revised equation, we double-checked formal descriptions and implementations of PoS coins. Table \ref{table: PoS comparison} shows leader selection processes and their consequent probabilities to be selected as a block proposer. 
In the table, \textsf{hash} denotes a one-way cryptographic hash function. Let $M_i$ be a miner $i$'s unique seed value, $T$ be a timestamp, and $D$ be a current difficulty in PoS. $C$ is the number of ages of staked coins by a miner. $A$ is a coin age. $p_i$ is the probability of being a leader for a miner $i$. $s_i$ is a miner $i$'s stake amount. $a_i$ is a miner $i$'s coin age.

\textbf{Peercoin} is the first cryptocurrency to adopt PoS. It applied PoW and PoS at the same time. It uses the concept of \textit{coin age} which is the time duration from the coin reception. A miner can check if the miner can be a block proposer every second by comparing a hash value named \textit{proofhashOfStake} \cite{PeercoinPoSCode} and $\textit{difficulty} \times \textit{stake amount} \times \textit{coin age}$. The hash value is the output of a cryptographic hash function with a miner's unique seed and the time value as inputs. The unique seed refers to former blocks and the miner's staking transactions. 
Therefore, it is hard to modify the seed value.
Assuming that every seed value is unique and unmodifiable, the cryptographic hash function outputs a random value from a uniform distribution. 
The probability to be a block proposer at every second is proportional to the miner's stake and coin age.

\textbf{Blackcoin} and \textbf{Nxt} are PoS cryptocurrencies mentioned in Saad et al. \cite{saad2021pos}. They are early PoS coins and followed Peercoin's PoS without the concept of \textit{coin age} because of security. Blackcoin \cite{blackcoin} described its PoS mechanism as $\textit{proofhash} < \textit{coins} \times \textit{target}$. The \textit{proofhash} corresponds to the hash value in Peercoin. The \textit{target} is the difficulty to control block generation speed. In the same way, if \textit{proofhash} is given randomly, the probability to be a proposer is proportional only to a miner's stake amount. It is similarly designed in Nxt \cite{NxtWhitepaper}.

\textbf{Ouroborous} and \textbf{Algorand} more directly select a block proposer. For example, Ouroboros \cite{kiayias2017ouroboros} has a special election function $\mathcal{F}$ which outputs a leader miner using a current stake distribution. It is designed to select a miner with a probability $p_i = \frac{s_i}{\sum^{n}_{k=1} s_k}$ \cite{kiayias2017ouroboros}. 
In Algorand, any miner can be a potential leader only with one token. Therefore, we assume that an adversary can split his stake into the least scale to be a potential leader. The adversary's probability is proportional to his stake ratio compared to the total stake amount. Therefore, the probability to be a next miner is $p_i = \frac{s_i}{\sum^{n}_{k=1} s_k}$, which is identical with Eq. \ref{equation: next block in real}.

\section{Conclusion}

In this comment paper, first, we provided a conventional concept of PoS based on real-world PoS cryptocurrencies. Second, we compared the probability to mint the next block in the conventional PoS with the probability presented in Saad et al. \cite{saad2021pos}.
Even though our conventional PoS concept covers Blackcoin and Nxt, which are mentioned in Saad et al., it does not match with the conventional PoS probability in \cite{saad2021pos}. It implies the fairness and decentralization issues that Saad et al. attempted to solve may not exist.

\bibliographystyle{IEEEtran}
\bibliography{references}

\end{document}